\documentclass[aps,pra]{revtex4}

\usepackage{epsfig,pstricks}
\usepackage{amsmath,amssymb}
\usepackage{color}

\def\CC{{\rm\kern.24em \vrule width.04em height1.46ex depth-.07ex
\kern-.30em C}}
\def\P{{\rm I\kern-.25em P}}
\def\NN{{\rm I\kern-.25em N}}
\def\RR{{\rm
         \vrule width.04em height1.58ex depth-.0ex
         \kern-.04em R}}
\def\id{{\rm 1\kern-.22em l}}
\def\ZZ{{\sf Z\kern-.44em Z}}

\newtheorem{pdef}{Definition}[section]
\newenvironment{defi}[1]{\begin{pdef} {\em  (#1)} \begin{quote}}{\end{quote}\end{pdef}}

\newtheorem{Theorem}{Theorem}[section]
\newenvironment{since}{${}^{}$\\ {\bf Proof}:  \begin{quote}\begin{em}}{\begin{flushright}
                   {\bf q.e.d.}\end{flushright}\end{em}\end{quote} }
\newenvironment{eqblock}[2]{\beq\label{#2}\begin{array}{#1}}{\end{array}
                                \eeq}
\newenvironment{neqblock}[1]{\[\begin{array}{#1}}{\end{array}\]}
\newcommand{\fat}[1]{\mbox{\boldmath $ #1 $\unboldmath}}
\newcommand{\beqb}{\begin{eqblock}}
\newcommand{\eeqb}{\end{eqblock}} 
\newcommand{\nbeqb}{\begin{neqblock}}
\newcommand{\neeqb}{\end{neqblock}} 
\newcommand{\eps}{\varepsilon}
\newcommand{\beq}{\begin{equation}}
\newcommand{\beqa}{\begin{eqnarray}}
\newcommand{\eeq}{\end{equation}}
\newcommand{\eeqa}{\end{eqnarray}}
\newcommand{\nbeqa}{\begin{eqnarray*}}
\newcommand{\neeqa}{\end{eqnarray*}}

\newcommand{\ket}[1]{| #1 \rangle}

\newcommand{\Matrix}[2]{\left( \begin{array}{#1} #2 \end{array}
  \right)}
\newcommand{\diag}{{\rm diag}\;}
\def\DJo{$\;$\kern-.4em \hbox{D\kern-.8em\raise.15ex\hbox{--}\kern.35em okovi\'c}}

\begin{document}

\title{Genuinely multipartite entangled states in higher dimensions: a generalization of balancedness}
\author{Andreas Osterloh}
\affiliation{Institut f\"ur Theoretische Physik, 
         Universit\"at Duisburg-Essen, D-47048 Duisburg, Germany.}
\email{andreas.osterloh@uni-due.de}
\begin{abstract}
I generalize the concept of balancedness to qudits with 
arbitrary dimension $d$. It is an extension of the concept of balancedness in 
New J. Phys. {\bf 12}, 075025 (2010) [1].
At first, I define maximally entangled states as being the stochastic states (with local reduced density matrices
$\id/d$ for a $d$-dimensional local Hilbert space) that are not product states
and show that every so-defined maximal genuinely multi-qudit
entangled state is balanced. Furthermore, all irreducibly balanced states are genuinely multi-qudit entangled
and are locally equivalent with respect to $SL(d)$ transformations (i.e. the local filtering transformations (LFO))
to a maximally entangled state. 
In particular the concept given here
gives the maximal genuinely multi-qudit entangled states for general local Hilbert space dimension $d$.
All genuinely multi-qudit entangled states are an element of the partly balanced $SU(d)$-orbits. 
\end{abstract}

\maketitle

\section{Introduction}

Entanglement is one of the cornerstones of quantum information theory and it gains 
an increasing importance in every branch of physics. In a work from 2000~\cite{MONOTONES} Vidal 
has listed the minimal requirements, a measure of entanglement has to satisfy, and
called such a quantity an entanglement monotone. In particular,
it has to be invariant under the local unitary group, and it must not increase under 
arbitrary local operations, considering also classical communications (LOCC). 
The transformations, entanglement has to be invariant with, has soon be enlarged to the stochastic version of 
LOCC, the SLOCC, and the relevant invariance group is the $SL$\cite{Duer00}, or the complex representations 
of the local unitary group\cite{ZeierComm}.
This has given rise to analyze the $SL$ invariant measures of entanglement more in detail
\cite{Coffman00,Wong00,Luque02,Miyake02,VerstraeteDM03,Miyake03,Miyake04,OS04,Luque05,OS05,DoOs08}.
Concerning the entangled states, it was observed immediately that they satisfy a ``center of mass'' 
condition for pure states of two, and three qubits\cite{Coffman00}.
This condition paved the way towards the balancedness condition, a condition that 
roughly states that the numbers of times the local state $\ket{0}$ appears, has to be equal to that 
of the occurrence of a local $\ket{1}$ on each qubit. 
It has been demonstrated to be due to the underlying $SL$ symmetry, and that every pure genuinely 
multi-qubit entangled state satisfies these conditions\cite{OS09}.
In the meantime a generalization of the method of local antilinear operators in Refs.~\onlinecite{OS04,OS05}
to qudits of dimensions $3$ and $4$ has appeared in Ref.~\onlinecite{Ost14} and the question of how 
this generalization would be transported into some condition for the genuinely multi-particle entangled states
was natural.
I introduce here a further generalization of the concept of balancedness to 
higher local dimensions. We will see that, roughly speaking, each of the numbers $0,\dots,2S$, 
which represent the $2S+1$ eigenstates of a local spin $S$, must occur the same number of 
times in order for a state being possibly maximally entangled. 
These are the states that are prescribed by the symmetry as genuinely many-qudit entangled
\footnote{There are two independent notions for the word 
``genuinely multipartite entanglement'' in the literature. 
The first goes back to the work
\cite{Bourennane04} and means {\em not bipartite}\/, 
whereas the notion used in this work follows \cite{OS05} and
means {\em not bipartite} and {\em distinguishable by a non-zero
 $SL(N)^{\otimes q}$-invariant}.
Non bipartite states of the null-cone, like the $W$-states, 
make the difference\cite{JohannsonO13}.}.

The outline of the article is as follows. After giving a brief introduction into the balancedness condition 
for many qubits in section \ref{Balanced}, a definition of maximally genuinely entanglement will be presented in Section 
\ref{Max}. 
Then, a generalization of the balancedness conditions is elaborated in Section \ref{States} for states of many qudits,
which are the generalized balanced states in this case.
The conclusions are drawn in Section \ref{Outlook}.

\section{Balancedness for qubit states}\label{Balanced}

It has been observed in Ref. \onlinecite{Coffman00} that for the three-tangle 
\beqa\label{def_3tangle}
\tau_3 &=& 4\ |d_1 - 2d_2 + 4d_3|\\
  d_1&=& \psi^2_{000}\psi^2_{111} + \psi^2_{001}\psi^2_{110} + \psi^2_{010}\psi^2_{101}+ \psi^2_{100}\psi^2_{011} \nonumber \\
  d_2&=& \psi_{000}\psi_{111}\psi_{011}\psi_{100} + \psi_{000}\psi_{111}\psi_{101}\psi_{010}\nonumber \\ 
    &&+ \psi_{000}\psi_{111}\psi_{110}\psi_{001} + \psi_{011}\psi_{100}\psi_{101}\psi_{010}\nonumber \\
    &&+ \psi_{011}\psi_{100}\psi_{110}\psi_{001} + \psi_{101}\psi_{010}\psi_{110}\psi_{001}\nonumber \\
  d_3&=& \psi_{000}\psi_{110}\psi_{101}\psi_{011} + \psi_{111}\psi_{001}\psi_{010}\psi_{100}\nonumber 
\eeqa
the components of the $d_i$ correspond to a line ($d_1$), a rectangle ($d_2$), or a tetrahedron ($d_3$)
with the property, that their ``center of mass'' coincides with that of the underlying cube. For two-qubit Bell states 
the same is valid for a two-dimensional square.
The generalization of this observation is called {\em balancedness}\cite{OS09}, 
and it stems from the local invariant operator, which is $\eps_{ij}$ in the case of qubits.
It is also known as the spinor metric. In other words, every spin-singlet
will be a candidate for a maximally genuinely many-qubit entangled state.
The simplest $SL(2)$ invariant two-qubit measure would then be the determinant of the state, 
when written in matrix form
$$
c(\psi)=2 (\psi_{00}\psi_{11}-\psi_{01}\psi_{10})=\eps_{ij}\eps_{kl}\psi_{ik}\psi_{jl}\; ,
$$
where we imposed Einstein sum convention.
Key ingredients are the binary matrix $\fat{B}$ or, equivalently, the alternating matrix $\fat{A}$.
For example, the state $\alpha_1 \ket{111}+\alpha_0 \ket{000}$ for $\alpha_i\neq 0$ has the binary matrix
$$
\fat{B}=\Matrix{cc}{1&0\\1&0\\1&0}\; ,
$$
and the corresponding alternating matrix is
$$
 \fat{A}=\Matrix{cc}{1&-1\\1&-1\\1&-1}\; .
$$ 
The balancedness condition is then written as
\beq
\sum_{l=1}^L A_{il}n_l =0
\eeq
where $n_l\in \NN_+$ and for all $i=1,\dots , q$, where $q$ is the number of qubits.
The number $l$ is running from $1$ to $L$, which is the length of the state.
Here, $L$ is to be understood as the minimum length the state can have on its local $SU(2)$ orbit, 
which is an invariant with respect to $SL(2)$ as well, and therefore
balancedness is also an $SU(2)$ invariant concept. We refer to Ref.~\onlinecite{OS09} for
the details, but we want to notice here that every state whose form of minimal length is balanced is 
genuinely multi-qubit entangled or a product of genuinely entangled $q_i$ qubit states ($q_i\geq 2$)
such that $\sum_i q_i = q$; 
every state whose form of minimal length is irreducibly balanced is genuinely multi-qubit entangled.
Furthermore, every maximal genuinely multi-qubit entangled state is balanced.
Also, every (irreducibly) balanced state
is proven to be $SL$ equivalent to a stochastic state.
This means that there are different classes of genuinely entangled states, depending whether 
they have a form of minimal length that is (irreducibly) balanced, or which contains still 
an unbalanced part - the partly (irreducibly) balanced states. The difference of the two is that 
whereas the (irreducibly) balanced states are $SL$ equivalent to a stochastic state, this is only true
for the partly (irreducibly) balanced states after an infinite sequence of local filtering operations. 

\section{Maximally entangled states}
\label{Max}

At the beginning we want to define what we consider a maximally entangled 
$q$ qudit state.
\begin{defi}{Maximal genuine multi-qudit entanglement}\label{MaxEnt}
We call a pure $q$-qudit state $\ket{\psi_q}$ {\em maximal genuine multipartite qudit entangled},
{\em iff}

\begin{enumerate}
\roman{enumi}
\item the state is not a product, i.e. the minimal rank of any reduced density matrix of $\ket{\psi_q}$ 
is $2$. 
\item it is stochastic\cite{VerstraeteDM03}, that is, it's reduced single qudit density matrices
are all equal to $\id_{d}/d$ for a qudit of dimension $d$.
\end{enumerate}
\end{defi} 
This definition is a direct generalization of the one used in Ref.~\onlinecite{OS09}
and is widely accepted in the community.
It follows the principle that a maximally entangled state has maximal local
uncertainty. 
We will demonstrate below that maximally and genuinely multi-qudit entanglement is furthermore 
directly connected to $SL$-invariance, as it turned out for qubits\cite{OS09}.
Please note that here, we do not consider entanglement of fewer than $q$ particles, as is the case 
for $W$ type of states for qubit systems, nor an entanglement that does not fill all 
the Hilbert space, in the sense that some local density matrix is $\id/{d'}$, with $d'<d$.  
These types of entanglement only range over part of the available resources.
These resources are
already classifiable by some corresponding $SL$-invariant measures of fewer particles, 
as e.g. the $W$ state in qubit systems is already classified by the concurrence.
In the second case, the entanglement can be classified by $SL$-invariants of 
a smaller Hilbert space dimension $d'<d$. 
  
\section{Maximally entangled states predicted by the symmetry}
\label{States}
We have the determinant of the two-qudit state when written in matrix form,
$\ket{\psi_q}=(\psi_q^{i,j})_{i,j}$, 
as the simplest polynomial $SL$-invariant measure\cite{Cereceda03,GourGconcurrence05,Ost14}, 
and all $SL$-invariant measures are constructed from its local invariant operators 
(see Ref. \onlinecite{Ost14} for dimensions $3$ and $4$)
that are connected to $\eps_{i_1,\dots,i_d}$.
This determinant is a combination of elements of a d-dimensional 
square matrix, such that each row and column occurs precisely once.
We want to stress here that this means that each element of the local Hilbert space, 
$\ket{0}$ to $\ket{2S}$, is occurring once. 
This will define the generalization of 
the term ``balancedness'' for qubits to an arbitrary
dimension of the local Hilbert space dimension, hence arbitrary qudits. 
We want to add that the balancedness condition for qubits~\cite{OS09}
also tells that each element, $\ket{1}$ and $\ket{0}$, is occurring with exactly the same multiplicity
for each qubit.
Hence, the corresponding logic applies for qudits as well as for qubits.
We now have to cast this observation into an equation or a set of 
equations, which will be the generalized balancedness conditions each genuinely multi-qudit entangled
state has to satisfy.

To this end, we consider a pure $q$ qudit state of local Hilbert space dimension
$(2S+1)$
\beq
\ket{\psi_q}=\sum_{i_1, \dots , i_q=1}^{d=2S+1} \psi_{i_1,\dots,i_q}\ket{i_1,\dots , i_q}\;
. 
\eeq
The state is represented in an orthogonal product basis.
What we ask for, is first an analogue of the binary matrix $B_{ik}$ from Ref.~\onlinecite{OS09}
where the weights are not written due to the underlying $SL(d)$.
We call the number of orthogonal 
product vectors the length $L$ of this representation. 
E.g. for the single site state of a qutrit ($S=1$) 
$\ket{\Psi}=\alpha_0\ket{0}+\alpha_1\ket{1}+\alpha_2\ket{2}$ 
this would mean, $B_{\ket{\Psi}}=(0,1,2)$, and length $L=3$.
For a 2 qutrit state 
$\ket{\Psi}=\alpha_0\ket{00}+\alpha_1\ket{11}+\alpha_2\ket{22}$ it takes the 
form
\beq\label{twoQtrit}
B_{\ket{\Psi}}=\Matrix{ccc}{0&1&2\\0&1&2}\; ,
\eeq
which has also length $L=3$.

For representing the condition that each basis state has to occur the same number of times,
note that the $n$-th roots of unity
sum up to zero, $\sum_{j=1}^n e^{i 2 \pi \frac{j}{n}}=0$. 
Therefore, one idea is to take a state in the local Hilbert space of 
spin $2S+1$, and attribute to it precisely one number out of $\{0,\dots,2S\}$.
If $2S+1$ is prime, this works without parts of the sum
yet being zero. 
This leads to the condition for spin $S$ ($2S+1$ prime, e.g. $S=1,2,3,5,6,\dots$ )
\beq\label{balanced}
\exists\, n_1,\dots,n_L\,\in\NN_{+} \ni \quad
\sum_{k=1}^L n_k e^{i 2 \pi \frac{B_{jk}}{2S+1}} =0
\; ; \ \forall \ j=1,\dots , q
\eeq
with $q$ being the number of qudits and $L$ being the length of the state. 
This condition defines generalized balancedness for $2S+1$ being prime.

At first we briefly show that for spin-1/2, the above exponential gives 
precisely the alternating matrix.
In this case the condition reads $\sum_{k=1}^L e^{i \pi B_{jk}} n_k=\sum_{k=1}^L A_{ik} n_k=0$.
The maximally entangled state for this case is\cite{OS09}
\beq\label{MAXENT}
\ket{\psi}=\sum_{k=1}^L \sqrt{n_k} {\prod_{j=1}^{q}}^\otimes\ket{B_{jk}}\; ,
\eeq
whose length is $L=\sum_k n_k$.
One could hence think of a matrix $A_{jk}$ which is
\beq
A_{jk}=e^{i 2 \pi \frac{B_{jk}}{2S+1}}\; .
\eeq

For $2S+1$ not being prime, one still has the $2S$ conditions that every 
two occurrences appear the same number of times. For this partial conditions
let e.g. $(A_{\ket{\psi}}^{(j,j+1)})_{kl}$ be the matrix whose entry is -1 (1) {\em iff} 
the $k$-th qubit of the $l$-th basis state in $\ket{\psi}$ is $j$ ($j+1$) and 
$0$ if it has some other value. Thus,
\beq
\exists \ n_k\in \NN_+\ \ni \quad
\sum_{k=1}^L n^{(j)}_k (A_{\ket{\psi}}^{(j,j+1)})_{lk} =0\qquad \forall\ l=1,\dots,q\;;\forall \ j=0,\dots ,2S\;.
\eeq
These are $2S$ conditions per qudit.
We want to outline that the $A_{\ket{\psi}}^{(j,j+1)}$ can be added up to yield $A_{\ket{\psi}}^{(j,j+s)}$
in the following way
\beqa
A_{\ket{\psi}}^{(j,j+1)}+A_{\ket{\psi}}^{(j+1,j+2)}&=&A_{\ket{\psi}}^{(j,j+2)}\; ,\\
A_{\ket{\psi}}^{(j,j+s)}+A_{\ket{\psi}}^{(j+s,j+s+1)}&=&A_{\ket{\psi}}^{(j,j+s+1)}\; .
\eeqa

To see what this means, we write down explicitly $(A_{\ket{\psi}}^{(j,j+1)})_{lk}$
for the maximally entangled two qutrit state from Eq.~\eqref{twoQtrit}
\beq
A_{\ket{\Psi}}^{(0,1)}=\Matrix{ccc}{-1 & 1 & 0\\-1& 1 & 0}
\ ; \ A_{\ket{\Psi}}^{(1,2)}=\Matrix{ccc}{0 & -1 & 1\\0& -1& 1}\; .
\eeq
This leads to an $n_k^{(j)}$ as follows
\beq
n^{(1)}=(1,1,m)\ ;\ n^{(2)}=(n,1,1)
\eeq
We therefore have to choose the free integers to be $m=n=1$ and hence
$n_k^{(j)}=n_k=(1,1,1)$ for $k=0,1$. 
Notice that on every qudit one can choose to make a 
basis change. Permutations are just examples for this change of basis.
Since the balancedness condition acts on every single qudit, this means that 
$n_k^{(j)}=n_k$ holds in general.\\
It must be stressed that an arbitrary solution for the $n_k$ with 
non-negative integer numbers can also be realized by writing the $k$th state
with a multiplicity of $n_k$, i.e. $n_k$ number of times. \\
Thus, we finally have
\beq\label{Balanced}
\exists \ n_k\in \NN_+\ \ni \quad
\sum_{k=1}^L n_k (A_{\ket{\psi}}^{(j,j+1)})_{lk} =0\qquad \forall\ l=1,\dots,q\;; \ \forall \ j=0,\dots ,2S\;.
\eeq
The corresponding genuinely multi-qudit entangled state again is
\beq
\ket{\psi}=\sum_{k=1}^L \sqrt{n_k} {\prod_{j=1}^{q}}^\otimes\ket{B_{jk}}\; ,
\eeq
The $n_k$ are chosen to be relatively prime.

This leads to the following definitions, which are directly transcribed from ~\cite{OS09}.
\begin{defi}{Balancedness}
\begin{enumerate}
\item A pure $SU$-orbit of a spin-S state is called balanced, 
{\em iff} the conditions \eqref{Balanced} are satisfied for every state in the orbit.
\item Let $\ket{\psi}$ be a balanced $SU$-orbit. It is called irreducibly balanced {\em iff}
for a state in its form of minimal length it cannot be split into different balanced parts.
\item An $SU$-orbit is called partly balanced if some (but not all) of the $n_k$ 
are admitted to be zero.
A partly balanced state is called reducible/irreducible {\em iff} its balanced part is reducible/irreducible.
\end{enumerate}
\end{defi}
We have to define also what we mean when calling an $SU$-orbit ``unbalanced''.
\begin{defi}{Complete unbalancedness}
We call a state out of an $SU$-orbit ``unbalanced'' if it is locally unitarily equivalent to 
a state without balanced part.
\end{defi}
The next theorem states the well-definedness of the concept.
\begin{Theorem}
Product states are not irreducibly balanced.
\end{Theorem}
In the proof for this theorem
the only modification is to replace $mn/2$ in \cite{OS09} by $mn/(2S+1)$. 
We hence do not show it here, and refer to \cite{OS09} instead.
\\
Also the next theorem translates directly to the higher dimensional version of spin-$S$. 
\begin{Theorem}
Every stochastic state (local reduced density matrices equal to $\id/(2S+1)$
for spin $S$) is balanced.
\end{Theorem}
\begin{since}
Consider a q-qudit state $\ket{\psi}$ that is stochastic. We can write the state
in the form
\beq
\Matrix{cccccccccc}{
w_1&\dots&w_m&w_{m+1}&\dots&w_{m+n}&\dots&w_{m+n+\dots+1}&\dots&w_{L-m-n-\dots}\\
0&\dots&0&1&\dots&1&\dots&2S&\dots&2S\\
\Phi_1&\dots&\Phi_m&\Phi_1^{'}&\dots&\Phi_m^{'}&\dots&\Phi_1^{(2S)}&\dots&\Phi^{(2S)}_{L-m-n-\dots}}
\; ,
\eeq
with weights $w_i\in\CC$.
Let some of the states out of $\Phi$, $\Phi^{'}$, \dots, $\Phi^{(2S)}$ coincide and call their superposition
$\Psi$, and their complements $\Psi_\perp$, $\Phi^{'}$, \dots, $\Phi^{(2S)}$. 
The state is then written as
$(\ket{0}+\alpha\ket{1}+\dots+\alpha^{(2S)}\ket{2S})\otimes\ket{\Psi} 
                + \beta\ket{0}\otimes\ket{\Psi_\perp} +\dots+
                  \beta^{(2S)}\ket{2S}\otimes\ket{\Phi^{(2S)}}$. Since the local density matrix 
is proportional to the identity matrix in every basis, we find that 
$\alpha=0$, \dots, $\alpha^{(2S)}=0$. Hence, the $\Phi_i$, \dots , $\Phi_i^{(2S)}$ 
are orthogonal in pairs. Defining $p_i=|w_i|^2$ we find
\beq\label{cond}
\rho^{(1)}=\Matrix{ccc}{
\sum_{i\in I_0^j} p_i &\dots&0\\
\vdots&\ddots&\vdots\\
0&\dots& \sum_{i\in I_{2S}^j} p_i }
=\frac{1}{2S+1}\id\; ,
\eeq
where $I_k^j$ are the column numbers such that the $j$-th qudit takes value $k=0,\dots,2S$.
These sums of $p_i$ have to be equal in pairs. The conditions we get out of \eqref{cond} are 
the same as that for balancedness.
\end{since}
For the maximal length of an irreducibly balanced state, we have
\begin{Theorem}
Every balanced q-qudit state of spin-S and length larger than $2Sq+1$ is reducible.
\end{Theorem}
\begin{since}
Balancedness means the existence of integers $n_k$, $k=1,\dots ,L$, such that
\eqref{Balanced} is satisfied.
Here we must demonstrate that every state of length $L>2Sq+1$ must be reducible if balanced.
Add a vertical cut ${\cal K}\cup {\cal K}'=L$ with 
$\left | {\cal K} \right |,\left | {\cal K}' \right |\geq \frac{2Sq+1}{2}$. Define 
$\vec{\alpha}^{{\cal K}}_j:=(\alpha^{{\cal K}}_{j;1},\dots,\alpha^{{\cal K}}_{j;q})$
such that 
$\alpha^{{\cal K}}_{j;l}=\sum_{k\in{\cal K}} n_k ( A_{\ket{\psi}}^{(j,j+1)})_{lk}$.
Irreducibility means that $\alpha^{{\cal K}}_{j;l}\neq 0$ for all $l=1,\dots ,q$
and some $j=1,\dots , 2S$.
Now make a cut $\kappa$ and $\kappa'$ in ${\cal K}$ and ${\cal K}'$ and define
$\tilde{\cal K}:=({\cal K}\setminus \kappa)\cup\kappa'$. 
Including arbitrary positive 
integers $m_{k}$, $k\in  \kappa'$, and keeping $m_{k}=n_k$ for 
$k\in {\cal K}$, 
one finds
\[
\sum_{k\in \tilde{\cal K}} m_{k} (A_{\ket{\psi}}^{(j,j+1)})_{ik}=
\alpha^{{\cal K}}_{j;i} - \sum_{k\in \kappa} m_{k} (A_{\ket{\psi}}^{(j,j+1)})_{ik}
+  \sum_{k\in \kappa'} m_{k} (A_{\ket{\psi}}^{(j,j+1)})_{ik}
\]
Irreducibility implies that for arbitrary such subsets ${\cal K}$ and $\kappa$ no integer numbers
$\tilde{m}_k\in\ZZ^{|\kappa|+|\kappa'|}$ do exist such that 
$\sum_{k\in \kappa \cup \kappa'}\tilde{m}_k (A_{\ket{\psi}}^{(j,j+1)})_{ik}=\alpha^{{\cal K}}_{j;i}$ for all 
$i\in\{1,\dots , q\}$
and all $j\in\{1,\dots, 2S\}$. 
Without loss of generality $(A_{\ket{\psi}}^{(j,j+1)})_{\kappa\cup \kappa'}$ has rank $q$ (due to a suitable choice 
of ${\cal K}$ and $\kappa$). This implies that each of the at most $2S q$ 
conditions can be satisfied, given that we can have up to $2S q$ variables.
\end{since}
Finally, we have a one-to-one correspondence of irreducibly balanced states 
and stochastic states, which have all single-qudit reduced density matrices proportional to the identity.
\begin{Theorem}
Every (irreducibly) balanced state of length $L\leq 2Sq+1$ is equivalent under local filtering operations 
$SL(d,\CC)^{\otimes q}$ to a stochastic state.
\end{Theorem}
\begin{since}
Let $a_j$, $j=1,\dots , L$ be the amplitudes of the product state written in the $j$-th
column of $B_\psi$.
Let the local filtering operations be
\beq\label{LFO}
{\cal T}_{LFO}^{(i)}=
\diag\{t_{0;i},\frac{t_{1;i}}{t_{0;i}},\dots,\frac{1}{t_{2S-1;i}}\}
\eeq
with $t_{k;i}=: t^{z_{k;i}}$ for some positive $t$, $z_{k;i}\in\CC$ and $\forall \ i=1,\dots , q$, 
and $\diag$ the diagonal matrix with given entries. It is to mention that the $k$-th entry
has exponent $z_{k;i}-z_{k-1;i}$, $z_{-1}:=z_{2S}:=0$, for every $k$.
We must demonstrate that after suitable such operations all the $a_j$ can be equalized.
Let the $n_k=1$ for all $k=1,\dots , L$ without loss of generality and 
$B_{i1}=0$ for all $i=1,\dots ,q$. 
Then, after the LFO we get
\[
a_jt^{\sum_i (z_{B_{ij};i}-z_{B_{ij}-1;i})}=:a_j \phi_j
\]
and a following division by $a_0\phi_0$ of the amplitudes, we get
\[
\frac{a_j}{a_0}t^{\sum_{i=1}^q (z_{B_{ij};i}-z_{B_{ij}-1;i}) -\sum_{i=1}^{q} z_{0;i}}\; .
\]
Equality for all these quantities therefore means
\[
\log_t \frac{a_0}{a_j}=\sum_{i=1}^q (z_{B_{ij};i}-z_{B_{ij}-1;i}-z_{0;i})\; .
\]
These are at most $2Sq$ independent variables due to the balancedness. 
Since we have 
$L-1$ conditions for the amplitudes to be satisfied, this results in a total length 
of at most $L=2Sq+1$
amplitudes for irreducibly balanced states which can be equalized. 
For $2S>q$ we obtain at most $2q^2$ independent variables and hence a 
maximal length of $L=2q^2+1<2Sq+1$.
The resulting state is stochastic.
\end{since}
It is important to notice that the above proof works as well, 
if the state is only reducibly balanced, and of length shorter than, 
or equal to, $2Sq +1$.

As the essence of this work, I will prove the following theorem, 
which states that every
irreducibly balanced states belongs to the class of semi-stable states, 
hence it is detected by some $SL$-invariant measures. A representative of 
the irreducibly balanced states, namely its normal form in the sense of
Ref. ~\cite{VerstraeteDM03} whose reduced single-site density matrix is $\id_{2S+1}/2S+1$,
is stochastic and hence is maximally entangled, according
to definition \ref{MaxEnt}.   
\begin{Theorem}
Every (irreducibly) balanced state is stable, hence it is not in the zero-class
with respect to SLOCC transformations. That means they are robust against 
infinitely many LFO's in $SL(2S+1)^{\otimes q}$ and possess a finite normal form
\cite{VerstraeteDM03}
\end{Theorem} 
\begin{since}
The state is assumed already in its normal form in that it is 
irreducibly balanced. Hence, the remaining class of LFO's is
given by Eq.~\eqref{LFO}. Therefore, each state from $\ket{0}$ until $\ket{2S}$
is occurring exactly the same number of times $p_i$ for each qudit 
$i\in\{0,\dots,q\}$. 
We assume that all the $n_k=1$ in Eq. ~\eqref{Balanced} at first. 
This leads to a factor of 
$$
1=
\left[ t_0 \cdot \frac{t_1}{t_0}  \cdots \frac{t_{2S-1}}{t_{2S-2}} \cdot 
       \frac{1}{t_{2S-1}} \right]^{p_i}
$$
distributed among the $L$ product states in the superposition.
In total, this amounts to a factor of
$$
1=\prod_{i=1}^q 
     \left[ t_0 \cdot \frac{t_1}{t_0}  \cdots \frac{t_{2S-1}}{t_{2S-2}} \cdot 
       \frac{1}{t_{2S-1}} \right]^{p_i}
$$
distributed among the $L$ product states in the superposition.
Now we assume all factors in front of the $(L-1)$ product states 
in the superposition to be smaller than one.
Then, the factor in front of the $L$'th product state will be larger 
than one, since the all over product is equal to one.\\
Now we let the $n_k\in \NN_+$, and relabel the product states in the 
superposition such that, without loss of generality, $n_L=\min\{n_k;k=1,\dots,L\}$. 
Then, the same line of thought applies also here, and we have that the
$L$'th product state is multiplied by a number greater than one.
\end{since} 
It is to be mentioned of course that the above proof applies to an arbitrary
balanced state, even if it is only partly balanced.
The form of the maximally entangled states is unchanged to the form
\eqref{MAXENT} for qubits.

\section{Conclusions}\label{Outlook}

The concept of generalized balancedness with respect to that for qubits 
in Ref.~\cite{OS09} is introduced
in terms of equal occurrence of each qudit basis state. It is an $SU$ invariant
concept, as for qubits, in that a state is called (irreducibly) balanced if it has in its 
$SU$ orbit a state of minimal length that is (irreducibly) balanced.
The well-definedness of this concept is demonstrated in showing that no 
product state can be an irreducibly balanced state. 
Based on the stochasticity of maximally entangled states\cite{VerstraeteDM03}
we manage to prove that stochasticity implies balancedness and that 
every (irreducible) balanced state with minimal length $L\leq 2Sq +1$ is 
$SL$-equivalent to a stochastic state. 
Finally it is proven that all states that are
(reducibly) balanced are robust against SLOCC transformations.
It must be highlighted that also those states which fall only partly 
into this scheme (hence having a (irreducibly) balanced and an unbalanced part) 
belong to the class of semi-stable states.
They are all detected by some $SL(d)$ invariant measure of entanglement in contrast to
the unstable states, that are in the $SL(d)$ null-cone.
The generalization is given for arbitrary local dimension.
The normal form of an arbitrary genuinely many-qudit entangled states with minimal length
is (irreducibly) balanced plus eventual some unbalanced part, which applies to qubits as well.

\section*{Acknowledgements}

I had fruitful discussions with R. Zeier,
and acknowledge financial support by the German Research Foundation within the SFB TR12.

%\bibliography{biblio,Qubits.SU} 

\end{document}